\documentclass[longauth, traditabstract]{aa} 
\usepackage{graphicx}
\usepackage{txfonts}
\usepackage{natbib}
\begin{document}

\title{{\em Herschel} observations of Extra-Ordinary Sources: 
Methanol as a probe of physical conditions in Orion~KL
\thanks{{\em Herschel} is an ESA space
    observatory with science instruments provided by European-led
    Principal Investigator consortia and with important participation
    from NASA.}}

\author{
S.~Wang,\inst{1}
E.~A.~Bergin,\inst{1}
N.~R.~Crockett,\inst{1}
P.~F.~Goldsmith,\inst{2}
D.~C.~Lis,\inst{3}
J.~C.~Pearson,\inst{2}
P.~Schilke,\inst{4,5}
T.~A.~Bell,\inst{3}
C.~Comito,\inst{4}
G.A. Blake\inst{6}
E.~Caux,\inst{7,8}
C.~Ceccarelli,\inst{9}
J.~Cernicharo,\inst{10}
F.~Daniel,\inst{10,11}
M.-L.~Dubernet,\inst{12,13}
M.~Emprechtinger,\inst{3}
P.~Encrenaz,\inst{11}
M.~Gerin,\inst{11}
T.~F.~Giesen,\inst{5}
J.~R.~Goicoechea,\inst{10}
H.~Gupta,\inst{2}
E.~Herbst,\inst{14}
C.~Joblin,\inst{7,8}
D.~Johnstone,\inst{15}
W.~D.  Langer\inst{2} 
W.~B. Latter\inst{16}
S.~D.~Lord,\inst{16}
S.~Maret,\inst{9}
P.~G.~Martin,\inst{17}
G.~J.~Melnick,\inst{18}
K.~M.~Menten,\inst{4}
P.~Morris,\inst{16}
H.~S.~P. M\"uller,\inst{5}
J.~A.~Murphy,\inst{19} 
D.~A.~Neufeld,\inst{20}
V.~Ossenkopf,\inst{13,21}
M.~P\'erault,\inst{11}
T.~G.~Phillips,\inst{3}
R.~Plume,\inst{22}
S.-L.~Qin,\inst{5}
S.~Schlemmer,\inst{5}
J.~Stutzki,\inst{5}
N.~Trappe,\inst{19}
F.~F.~S.~van der Tak,\inst{21}
C.~Vastel,\inst{7,8}
H.~W.~Yorke,\inst{2}
S.~Yu,\inst{2}
\and
J.~Zmuidzinas,\inst{3}
}
\institute{Department of Astronomy, University of Michigan, 500 Church Street, Ann Arbor, MI 48109, USA \\
\email{shiya@umich.edu}
\and Jet Propulsion Laboratory,  Caltech, Pasadena, CA 91109, USA
\and California Institute of Technology, Cahill Center for Astronomy and Astrophysics 301-17, Pasadena, CA 91125 USA
\and Max-Planck-Institut f\"ur Radioastronomie, Auf dem H\"ugel 69, 53121 Bonn, Germany
\and I. Physikalisches Institut, Universit\"at zu K\"oln,
              Z\"ulpicher Str. 77, 50937 K\"oln, Germany
\and California Institute of Technology, Division of Geological and Planetary Sciences, MS 150-21, Pasadena, CA 91125, USA
\and Centre d'\'etude Spatiale des Rayonnements, Universit\'e de Toulouse [UPS], 31062 Toulouse Cedex 9, France
\and CNRS/INSU, UMR 5187, 9 avenue du Colonel Roche, 31028 Toulouse Cedex 4, France
\and Laboratoire d'Astrophysique de l'Observatoire de Grenoble, 
BP 53, 38041 Grenoble, Cedex 9, France.
\and Centro de Astrobiolog\'ia (CSIC/INTA), Laboratiorio de Astrof\'isica Molecular, Ctra. de Torrej\'on a Ajalvir, km 4
28850, Torrej\'on de Ardoz, Madrid, Spain
\and LERMA, CNRS UMR8112, Observatoire de Paris and \'Ecole Normale Sup\'erieure, 24 Rue Lhomond, 75231 Paris Cedex 05, France
\and LPMAA, UMR7092, Universit\'e Pierre et Marie Curie,  Paris, France
\and  LUTH, UMR8102, Observatoire de Paris, Meudon, France
\and Departments of Physics, Astronomy and Chemistry, Ohio State University, Columbus, OH 43210, USA
\and National Research Council Canada, Herzberg Institute of Astrophysics, 5071 West Saanich Road, Victoria, BC V9E 2E7, Canada 
\and Infrared Processing and Analysis Center, California Institute of Technology, MS 100-22, Pasadena, CA 91125
\and Canadian Institute for Theoretical Astrophysics, University of Toronto, 60 St George St, Toronto, ON M5S 3H8, Canada
\and Harvard-Smithsonian Center for Astrophysics, 60 Garden Street, Cambridge MA 02138, USA
\and  National University of Ireland Maynooth. Ireland
\and  Department of Physics and Astronomy, Johns Hopkins University, 3400 North Charles Street, Baltimore, MD 21218, USA
\and SRON Netherlands Institute for Space Research, PO Box 800, 9700 AV, Groningen, The Netherlands
\and Department of Physics and Astronomy, University of Calgary, 2500
University Drive NW, Calgary, AB T2N 1N4, Canada
}


\abstract{
We have examined methanol emission from Orion KL  with of the {\em Herschel}/HIFI instrument, 
and detected two methanol bands centered at 524 GHz and 1061 GHz.
The 524 GHz methanol band (observed in HIFI band 1a) is dominated by the isolated $\Delta$J$=$0, K$=$-4$\rightarrow$-3, v$_t$$=$0
Q branch, and includes 25 E-type and 2 A-type transitions.
The 1061 GHz methanol band (observed in HIFI band 4b) is dominated by the $\Delta$J$=$0, K$=$7$\rightarrow$6, v$_t$$=$0
Q branch transitions which are mostly blended. 
We have used the isolated E-type v$_t$$=$0 methanol transitions to explore the physical conditions
in the molecular gas. 
With HIFI's high velocity resolution, the methanol emission contributed by different spatial components along the line of sight 
toward Orion KL (hot core, low velocity flow, and compact ridge) can be distinguished and studied separately.
The isolated transitions detected in these bands  cover a broad energy range (upper state energy ranging from 80 K to 900 K), 
which provides a unique probe of the thermal structure in each spatial component.  
The observations further show that the compact ridge is externally heated.
These observations demonstrate the power of methanol lines as probes of the physical conditions 
in warm regions in close proximity to young stars.
}

   \titlerunning{Methanol in Orion KL}
	\authorrunning{Wang et al.}
   \maketitle
%

\section{Introduction}

Methanol (CH$_3$OH) is an abundant molecule in a wide range of  interstellar conditions 
\citep[e.g.,][]{Friberg1988,Menten1988}. 
An asymmetric-top rotor, methanol has numerous rotational transitions 
from far-infrared to millimeter wavelengths.
This marks methanol as a common ``weed'' molecule,
and the number and strength of its transitions means that methanol contamination has to be minimized when analyzing
transitions from other molecules.
However, rather than being treated solely as a contaminant,
methanol is also a useful molecular tracer of dense gas \citep[][]{Menten1988}.
\citet{Leurini2004} further explore the properties of methanol emission in dense
molecular clouds and illustrate how methanol transitions at sub-millimeter and millimeter wavelengths
are sensitive to the density and the kinetic temperature of the gas.
Therefore, with its observational advantage of
a large number of lines being observable simultaneously,
methanol is an excellent tool with which to probe the physical structure
in dense molecular gas \citep[also see][]{Kama2010}.

A full spectral scan toward Orion KL covering frequency ranges 479.8 to 560.0 GHz (band 1a) 
and 1047.0 to1121.5 GHz (band 4b) 
has been carried out by the HIFI instrument \citep[][]{degrauw10}
aboard on the Herschel Space Observatory \citep[][]{pilbratt10},
as part of the Guaranteed Time Key Program {\em Herschel/HIFI Observations of Extraordinary
Sources: The Orion and Sagittarius B2 Star-forming Regions} (HEXOS).
Orion KL, located within the Orion molecular cloud at $\sim$ 450 pc, 
is the nearest massive star-forming region \citep[][]{Genzel1989}. 
It contains several kinematic components \citep[e.g.,][]{Blake1987,Persson2007},
a hot core, multiple outflows, and the compact ridge, surrounding by cold quiescent gas.
We have detected two methanol bands (i.e., groups of methanol lines within
small frequency ranges) with transitions
spanning a large range in upper state energy (80-900 K) falling within a single HIFI bandpass (4GHz).
In this paper, we show that these two bands of methanol can be used to determine
the gas temperature and we explore thermal gradients in the Orion compact ridge.
In \S 2, we discuss the HEXOS observations and the analysis tools used in this study.
In \S 3, we present the spectra of two methanol bands and their $^{13}$CH$_3$OH
counterparts, and resolve the methanol emission from the compact ridge and the outflow. 
In \S 4, we show the population diagrams for the detected isolated methanol transitions
for compact ridge and the outflow.
And in \S 5, we examine how these two methanol bands can be used to provide additional information
on the physical structure of the emitting regions.


\section{Observations}

\begin{figure*}
  \sidecaption
     \includegraphics[width=12cm]{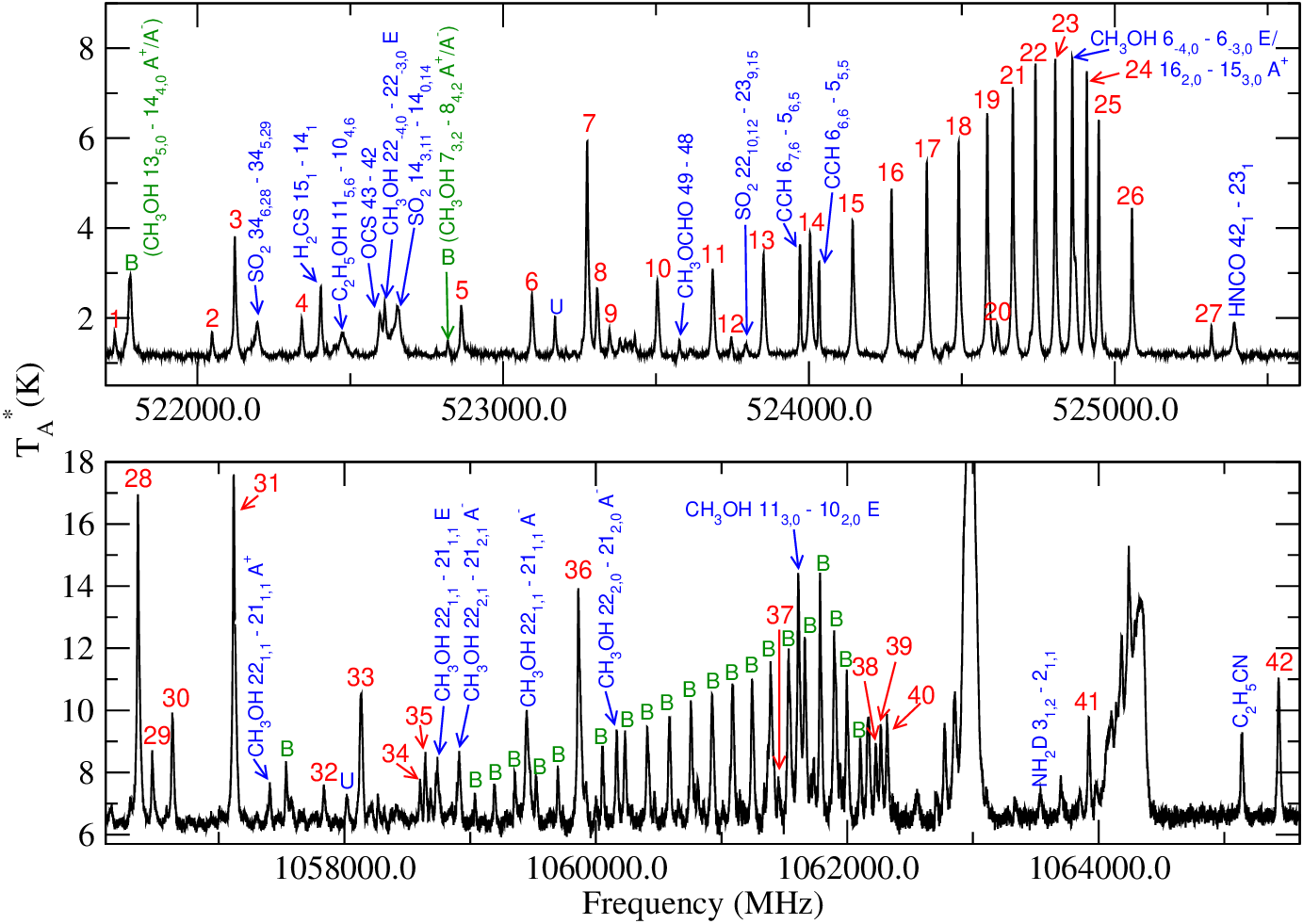}
   \caption{HIFI spectra of two methanol bands in band 1a (521.7-525.6 GHz; upper panel)
      and 4b (1056.1-1065.6 GHz; lower panel). Red number 1-42 label isolated methanol transitions,
      which are also listed in Table 1. Blue texts label transitions from other molecules
      and the methanol transitions which are blended. Green texts and "B" label the methanol lines
      which are blended with different parity states. ``U'' labels two unidentified lines.
     }
         \label{fig1}
\end{figure*}

The HIFI observations presented here, which are part of HEXOS \citep[][]{Bergin2010}, 
are full spectral scans in bands 1a and 4b, covering a frequency range
of 479.8 to 560.0 GHz and 1047.0 to 1121.5 GHz, respectively, pointing towards the Orion Hot
Core at $\alpha_{J2000} = 5^h35^m14.3^s$ and $\delta_{J2000} = -5^{\circ}22'36.7''$.
These observations were carried out 
on March 1 and 3, 2010 using the dual beam switch (DBS) mode with redundancy 6,
with the Wide Band Spectrometer (WBS) providing a spectral resolution of 1.1 MHz
over a 4~GHz IF bandwidth.
The HIFI beam sizes are $\sim$ 43$''$ in band 1a and 20$''$ in band 4b, and the beam efficiency in
both bands is 0.71. 
All data presented here were processed with HIPE \citep[][]{Ott2010} version 2.4,
using the standard HIFI deconvolution ($doDeconvolution$ task in HIPE).
From line-free regions in the spectrum, we obtain a rms of T$_A$$^*$$=$0.065 K for band 1a
and T$_A$$^*$$=$0.116 K for band 4b.
All spectra here were analyzed using the XCLASS program 
(https://www.astro.uni-koeln.de/projects/schilke/XCLASS), which provides access to the
CDMS and JPL molecular catalogs \citep[][]{CDMS_1,CDMS_2,JPL-catalog} for identifying lines
and obtaining the molecular parameters of methanol \citep[][]{Xu1997,Xu2008}.

\section{Results}

\subsection{Detection of distinctive methanol bands}

There are 223 and 141 methanol lines with v$_t$$=$0 and 1 detected in HIFI bands 1a and 4b, respectively.
No molecular and frequency information for transitions with v$_t$$>$1 is available at present from the CDMS and JPL catalogs.
Among these detected methanol lines, 112 in band 1a and 73 in band 4b are isolated transitions. 
The rest are transitions which are blended with either other methanol transitions or lines of other molecules.
These detected isolated methanol transitions span a wide range in upper state energy (0-1000 K).
The focus of this paper is to investigate how we can
utilize the advantage offered by methanol --  having many transitions within small frequency ranges --
to probe the physical conditions in the emitting region.
For this purpose, we  look for groups of isolated methanol transitions which are confined within 
small frequency range in the spectrum but which also cover a wide range in upper state energy
with minimal line blending.
This approach is useful as it shows an example of how physical conditions can be probed by
observing only a limited spectral range.

Among all the detected methanol lines, there are two very distinctive methanol bands in HIFI bands 1a and 4b-- 
both are constructed with a series of methanol lines confined within a few GHz range
and have a similar characteristic spectral shape.   
Fig.~\ref{fig1} shows the spectra of these two methanol bands.
The first covers 521.7 to  525.6 GHz in HIFI band 1a (upper panel).  
The 4 GHz spectrum includes the methanol band peaked at 524.8 GHz, which is
defined as the ``524 GHz methanol band'' in this paper.
The second covers from 1056.1 to 1065.6 GHz in HIFI band 4b (lower panel).  
The 10 GHz spectrum includes the methanol band 
peaked at 1061.6 GHz, which is defined as the ``1061 GHz methanol band'' in this paper.

In the 524 GHz methanol band, 
31 methanol lines are detected: 27 isolated methanol transitions (labeled in red numbers 1-27),
2 lines either blended with other methanol transitions or transitions from other molecules
(labeled in blue), and 2 lines which are blended with methanol transitions having the same 
quantum numbers (J, K, v$_t$) but different parity with either E, A$+$, or A$-$ state 
(defined as the blended ``B'' lines and labeled as ``B'' in green). 
Methanol has E and A symmetric states, which are
considered as different species as no radiative transitions are allowed between them. 
A-type methanol can be further separated as $+$ and $-$ parity.
Fig.~\ref{fig1} shows that the 524 GHz methanol band is dominated by isolated transitions,
which are mostly $\Delta$J$=$0, K$=$-4$\rightarrow$-3 Q branch with v$_t$$=$0
representing the torsional ground state.
On the other hand, in the 1061 GHz methanol band, 42 methanol lines are detected: 
15 isolated transitions (labeled in red numbers 28-42), 
6 lines blended with others, and 21 blended ``B'' lines. 
Fig.~\ref{fig1} shows that, unlike the 524 GHz methanol band,
the 1061 GHz methanol band is dominated by the blended ``B'' lines mostly
with $\Delta$J$=$0, K$=$7$\rightarrow$6, v$_t$$=$0 transitions.

\begin{figure}
   \centering
   \includegraphics[width=0.9\columnwidth]{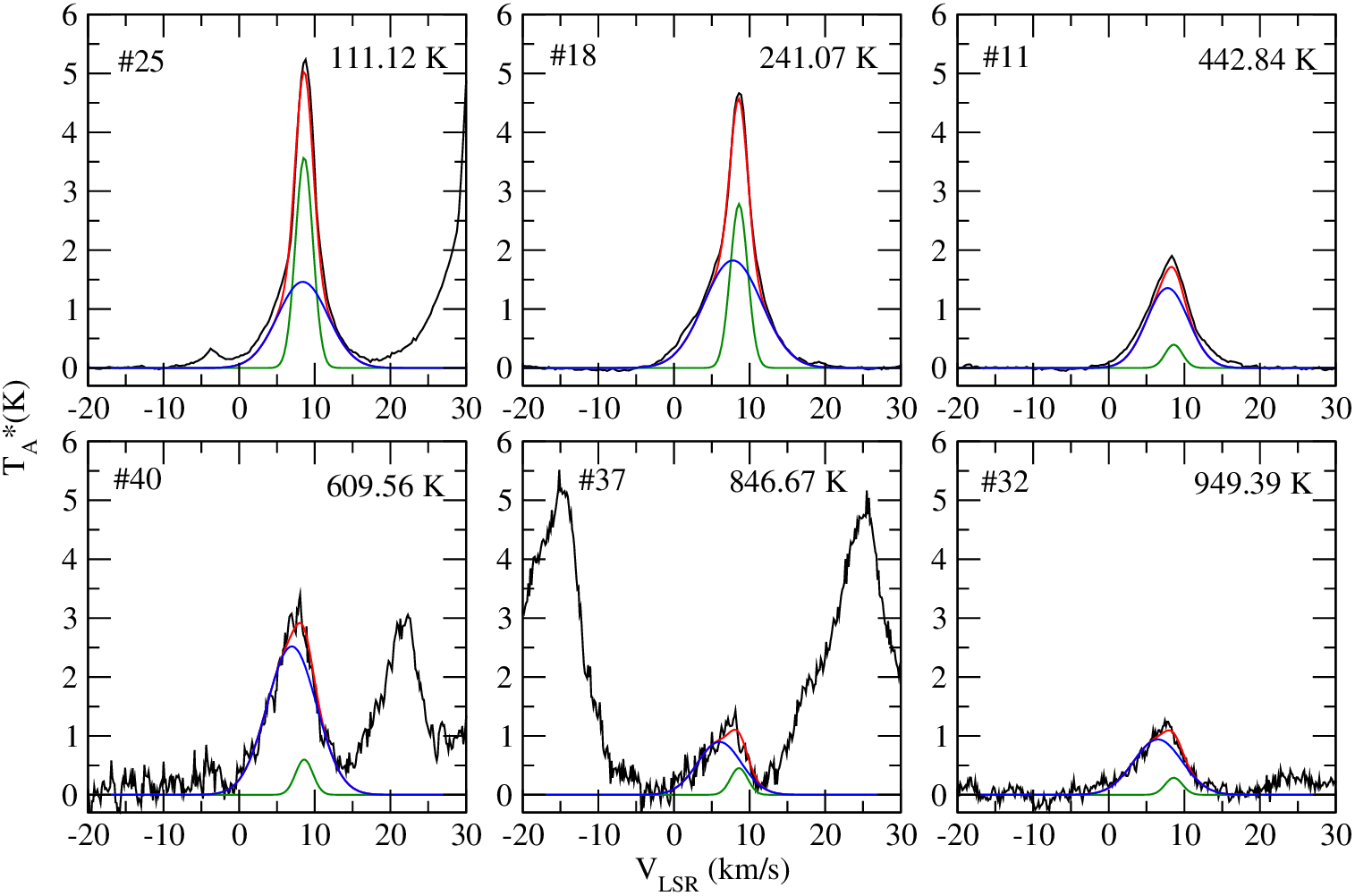}
   \caption{
     Spectra of six isolated methanol lines. Each spectrum is plotted with the data
     (black) and the two-component Gaussian fit (red) to the data, which includes a
     narrower component (green) and wider component (blue). The number in the upper left
     corner and the energy (in K) in the upper right corner
     of each panel indicate the associated methanol transition and its upper state energy
   given in Table 1.       
     }
         \label{fig2}
\end{figure}

Table 1 lists the molecular parameters of the 42 isolated transitions,
including 37 E-type (33 v$_t$$=$0 and 4 v$_t$$=$1) and 5 A-type 
(4 v$_t$$=$0 and 1 v$_t$$=$1) methanol transitions,
their peak intensities in units of antenna temperature (T$_A$$^*$), line frequency $\nu$, 
upper state energy E$_u$, and S$\mu^2$ where S and $\mu$ are the line strength and dipole moment, respectively. The peak intensities are estimated with
baseline continuum subtracted;  the continuum level $\sim$ 1.23 K for band 1a  and
$\sim$ 6.45 K for band. 
These methanol lines have peak intensities between 0.4 and 12 K.  They are widely spread in upper state energy, which ranges
from 80 to 900 K, covering almost the same energy range as those detected from the entirety of methanol lines detected in bands 1a and 4b.
This energy range is sensitive to kinetic temperatures anticipated in molecular clouds, and
these lines are thus a good set of transitions to probe the cloud conditions. 
We only focus on those 33 isolated E-type
v$_t$$=$0 (25 in band 1a and 8 in band 4b), in order not to get too  involved in complicated excitation issues,
any potential A/E abundance variations, and any v$_t$$>$0 transitions that can be  excited through radiative
pumping of the strong N$\nu_{12}$ torsional bands.

Furthermore, among all 112 isolated methanol lines detected in the whole of band 1a, there are in total
55 isolated E-type v$_t$$=$0 transitions. This means that 25 (almost 50\%) out of the total 55 lines
in band 1a (covering 80 GHz) are located within the 4 GHz bandwidth of this 524 methanol band. 
On the other hand, among the 73 isolated methanol lines detected in whole band 4b,
there are in total 36 isolated E-type v$_t$$=$0 transitions-- only $\sim$ 20\% of them are located in the
1061 GHz methanol band. 
The fact that the 524 GHz methanol band is dominated by isolated
transitions while the 1061 GHz methanol band is mostly constructed with blended ``B'' lines,
suggests that the 524 GHz methanol band is  more suitable for exploration of the physical conditions.
Hereafter, when we use the terms  ``524 GHz methanol band'' and ``1061 GHz methanol band'' 
we are referring only to those isolated E-type v$_t$$=$0 transitions.

\begin{figure}
   \centering
   \includegraphics[width=0.75\columnwidth]{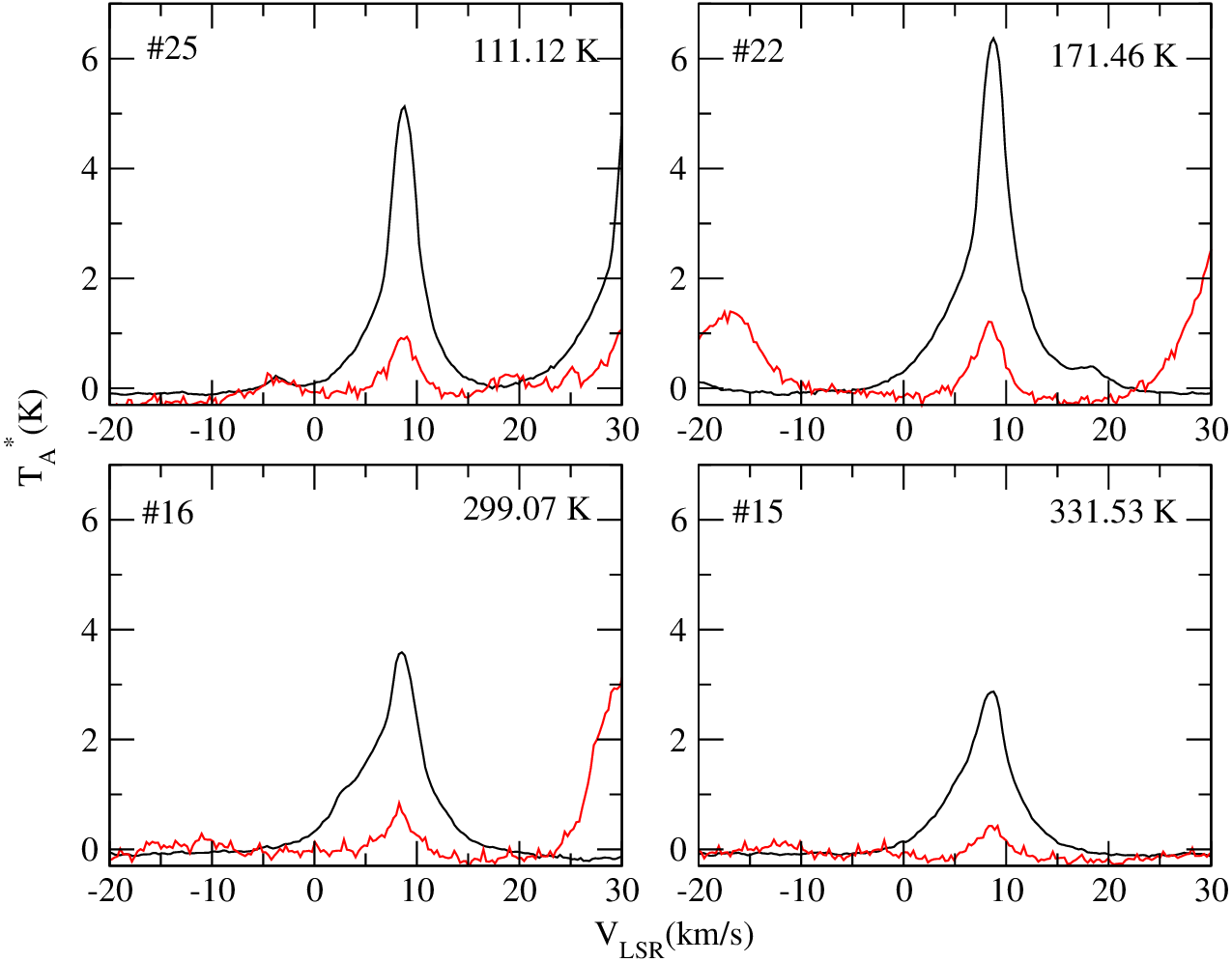}
   \caption{
     Spectra of four isolated methanol lines (black) plotted with their $^{13}$CH$_3$OH counterparts (red),
     which are scaled by a factor of two in order to be compared to methanol lines more clearly. 
     The number in the upper left
     corner and the energy (in K) in the upper right corner
     of each panel indicate the associated methanol transition and its upper state energy
     given in Table 1.   
     }
         \label{13ch3ohfig}
\end{figure}

\begin{table}
\caption{Detected isolated CH$_3$OH Lines}
\centering
\begin{tabular}{cccccc}
\hline
label & CH$_3$OH & $\nu$ & E$_u$ & S$\mu^2$ & T$_{A}$$^*$ \\
 & (J$_{K,{\rm{v}}_t}$) & (MHz) & (K) & (D$^2$) & (K) \\
\hline
1 & 25$_{-4,0}$$\rightarrow$25$_{-3,0}$ E & 521728.127 & 841.28 & 22.047 & 0.44 \\    
2 & 24$_{-4,0}$$\rightarrow$24$_{-3,0}$ E & 522046.360 & 783.41 & 21.209 & 0.45 \\
3 & 9$_{4,0}$$\rightarrow$10$_{3,0}$ E    & 522121.802 & 200.44 & 1.997 & 2.58 \\   
4 & 23$_{-4,0}$$\rightarrow$23$_{-3,0}$ E & 522340.009 & 727.83 & 20.360 & 0.74 \\
5 & 21$_{-4,0}$$\rightarrow$21$_{-3,0}$ E & 522861.608 & 623.60 & 18.646 & 1.05 \\
6 & 20$_{-4,0}$$\rightarrow$20$_{-3,0}$ E & 523092.856 & 574.94 & 17.773 & 1.28 \\
7 & 14$_{-1,0}$$\rightarrow$13$_{0,0}$ E & 523274.293 & 248.94 & 10.900 & 4.70 \\
8 & 19$_{-4,0}$$\rightarrow$19$_{-3,0}$ E & 523306.240 & 528.60 & 16.897 & 1.44 \\
9 & 24$_{1,0}$$\rightarrow$23$_{2,0}$ A$^+$ & 523346.243 & 703.52 & 10.320 & 0.51 \\
10& 18$_{-4,0}$$\rightarrow$18$_{-3,0}$ E & 523502.932 & 484.56 & 16.008 & 1.59 \\
11& 17$_{-4,0}$$\rightarrow$17$_{-3,0}$ E & 523683.978 & 442.84 & 15.112 & 1.86 \\
12& 24$_{3,0}$$\rightarrow$23$_{4,0}$ A$^-$ & 523744.752 & 745.74 & 7.788 & 0.35 \\
13& 16$_{-4,0}$$\rightarrow$16$_{-3,0}$ E & 523850.320 & 403.43 & 14.206 & 2.18 \\
14& 15$_{-4,0}$$\rightarrow$15$_{-3,0}$ E & 524002.815 & 366.32 & 13.290 & 2.64 \\
15& 14$_{-4,0}$$\rightarrow$14$_{-3,0}$ E & 524142.239 & 331.53 & 12.360 & 2.92 \\
16& 13$_{-4,0}$$\rightarrow$13$_{-3,0}$ E & 524269.306 & 299.07 & 11.423 & 3.64 \\
17& 12$_{-4,0}$$\rightarrow$12$_{-3,0}$ E & 524384.667 & 268.91 & 10.468 & 4.20 \\
18& 11$_{-4,0}$$\rightarrow$11$_{-3,0}$ E & 524488.918 & 241.07 & 9.499 & 4.67 \\
19& 10$_{-4,0}$$\rightarrow$10$_{-3,0}$ E & 524582.604 & 215.55 & 8.511 & 5.32 \\
20& 24$_{-2,0}$$\rightarrow$24$_{-1,0}$ E & 524614.631 & 724.08 & 20.883 & 0.62 \\
21& 9$_{-4,0}$$\rightarrow$9$_{-3,0}$ E & 524666.219 & 192.34 & 7.501 & 5.90 \\
22& 8$_{-4,0}$$\rightarrow$8$_{-3,0}$ E & 524740.210 & 171.46 & 6.461 & 6.44 \\   
23& 7$_{-4,0}$$\rightarrow$7$_{-3,0}$ E & 524804.977 & 152.90 & 5.380 & 6.51 \\   
24& 5$_{-4,0}$$\rightarrow$5$_{-3,0}$ E & 524908.214 & 122.73 & 3.018 & 6.26 \\   
25& 4$_{-4,0}$$\rightarrow$4$_{-3,0}$ E & 524947.258 & 111.12 & 1.647 & 5.18 \\    
26& 4$_{-3,0}$$\rightarrow$5$_{-2,0}$ E & 525055.729 & 85.92 & 0.560 & 3.21 \\  
27& 13$_{1,0}$$\rightarrow$12$_{-2,0}$ E & 525315.119 & 232.30 & 0.169 & 0.57 \\   
28& 11$_{-4,0}$$\rightarrow$10$_{-3,0}$ E & 1056354.886 & 241.08 & 8.659 & 10.58 \\
29& 18$_{0,1}$$\rightarrow$17$_{1,1}$ E & 1056466.663 & 696.17 & 8.313 & 2.32 \\
30& 22$_{-1,0}$$\rightarrow$21$_{-1,0}$ E & 1056627.148 & 590.67 & 17.746 & 3.53 \\
31& 6$_{-3,0}$$\rightarrow$5$_{-2,0}$ E & 1057117.722 & 111.46 & 5.541 & 11.21 \\
32& 22$_{-2,1}$$\rightarrow$21$_{-2,1}$ E & 1057831.005 & 949.39 & 17.720 & 1.19 \\
33& 9$_{-7,1}$$\rightarrow$8$_{-6,1}$ E & 1058132.994 & 627.54 & 12.064 & 4.12 \\
34& 22$_{2,1}$$\rightarrow$21$_{2,1}$ A$^+$ & 1058598.551 & 883.67 & 17.603 & 1.37 \\
35& 22$_{0,1}$$\rightarrow$21$_{0,1}$ E & 1058639.968 & 885.60 & 17.797 & 2.23 \\
36& 13$_{-2,0}$$\rightarrow$12$_{-1,0}$ E & 1059858.944 & 237.30 & 6.157 & 7.47 \\
37& 22$_{-7,0}$$\rightarrow$21$_{-7,0}$ E & 1061448.476 & 846.67 & 16.095 & 1.37 \\
38& 22$_{-5,0}$$\rightarrow$21$_{-5,0}$ E & 1062221.782 & 710.72 & 16.893 & 2.44 \\
39& 22$_{3,0}$$\rightarrow$21$_{3,0}$ A$^+$ & 1062261.800 & 636.61 & 17.458 & 3.06 \\
40& 22$_{2,0}$$\rightarrow$21$_{2,0}$ E & 1062312.231 & 609.56 & 17.571 & 3.41 \\
41& 22$_{2,0}$$\rightarrow$21$_{2,0}$ A$^+$ & 1063916.882 & 625.04 & 17.739 & 3.27 \\
42& 18$_{1,0}$$\rightarrow$17$_{0,0}$ E & 1065427.650 & 417.93 & 7.792 & 4.51 \\
\hline
\end{tabular}
\end{table}

\subsection{Resolving different kinematic components in Orion KL}

\begin{figure*}
  \sidecaption
     \includegraphics[width=12cm]{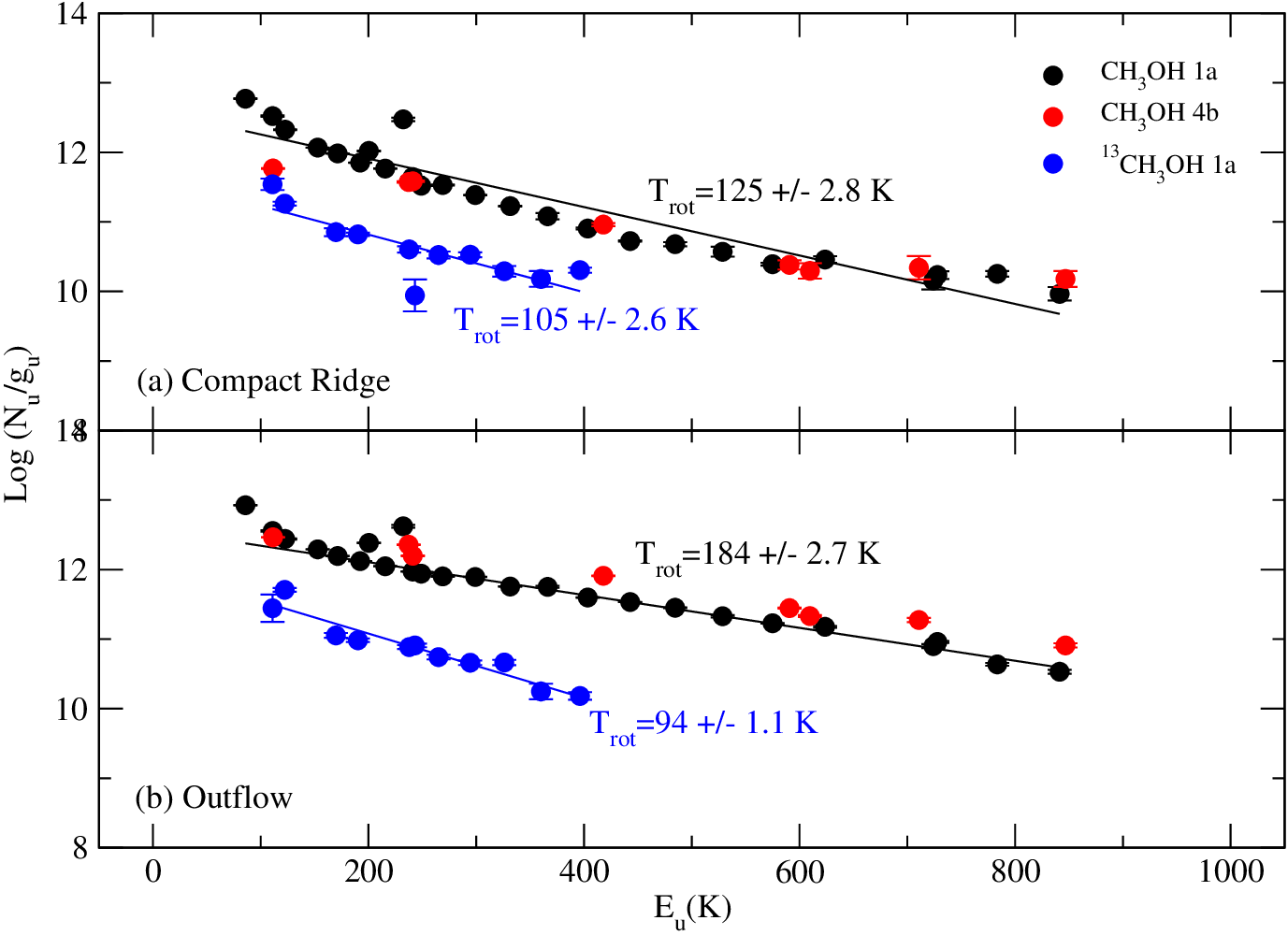}
   \caption{
     Population diagrams of detected 33 E-type v$_t$$=$0 methanol transitions for 
     (a) the compact ridge and (b) the outflow. Transitions in HIFI bands 1a and 4b
     are plotted in black and red, respectively. Transitions of  $^{13}$CH$_3$OH
     are plotted in blue. Straight lines are fitted in order to obtain temperature estimates
     assuming LTE.
     }
         \label{fig3}
\end{figure*}

Fig.~\ref{fig2} plots the spectra of six isolated methanol lines
with different upper state energies. All methanol lines show asymmetry in the line profiles,
indicating the existence of multiple contributions from different kinematic components in Orion KL,
as suggested previously.
For example, a high resolution VLA spectrum of one methanol line toward Orion KL \citep[][]{Wilson1989}
shows one narrower (linewidth $\Delta$V$\sim$2 km s$^{-1}$, and V$_{LSR}$$=$7.6 km s$^{-1}$)
and one wider ($\Delta$V$\sim$7 km s$^{-1}$, and V$_{LSR}$$\sim$6.2-9.2 km s$^{-1}$) component.
Consistent with previous results, the detected isolated methanol lines in Table 1 
can all be fit with two Gaussians (see Fig.~\ref{fig2}):
(1) a narrow component at V$_{LSR}$$=$8.6 km s$^{-1}$ with $\Delta$V$=$2.7 km s$^{-1}$,
and (2) one wide component at V$_{LSR}$$=$6-9 km s$^{-1}$ with $\Delta$V$=$4-11 km s$^{-1}$. 
Therefore, we fitted all methanol lines with fixed V$_{LSR}$ at 8.6 km s$^{-1}$ and
$\Delta$V at 2.7 km s$^{-1}$ for the narrow component, and freely fitting the V$_{LSR}$
and $\Delta$V for the wide component.
It is well known that there are distinct spatial components (hot core, outflow, and compact
ridge) along the line of sight that can be distinguished by their kinematic signature
in the molecular emission profile \citep[e.g.,][]{Blake1987,Persson2007}.
Based on this information, we assign the V$_{LSR}$$=$8.6 km s$^{-1}$ component to the compact ridge,
and the V$_{LSR}$$=$6-9 km s$^{-1}$ component to the outflow.

\subsection{$^{13}$CH$_3$OH counterparts for those detected methanol transitions}

In order to estimate the opacities of these $^{12}$CH$_3$OH transitions, we look for their 
$^{13}$CH$_3$OH counterparts throughout the whole band 1a.
13 of these 25 methanol transitions in HIFI band 1a with E$_u$ $<$ 370 K have $^{13}$CH$_3$OH counterparts detected
(Table 2). We did not search $^{13}$CH$_3$OH for those transitions in 4b because no molecular information
of $^{13}$CH$_3$OH transitions with frequency above 1 THz are available in either the CDMS and JPL catalogs. 
Fig.~\ref{13ch3ohfig} plots four selected methanol lines and their $^{13}$CH$_3$OH counterparts.
Following the same strategy to separate different kinematic components as shown in \S 3.2,
we then obtain the spectra for the compact ridge and the outflow component of each $^{13}$CH$_3$OH line. 
By comparing CH$_3$OH and $^{13}$CH$_3$OH lines, optical depth ($\tau$) 
for compact ridge and outflow components can be estimated using
\begin{equation}
\frac{T_{12}}{T_{13}} = \frac{T_{ex,12}(1-e^{-\tau_{12}})}{T_{ex,13}(1-e^{-\tau_{13}})},
\end{equation}
where we have assumed that  CH$_3$OH and $^{13}$CH$_3$OH have  the same excitation temperature,
CH$_3$OH is optically thick, and $\tau_{13}$$=$$\tau_{12}$/f, where f is the 13C to 12C abundance ratio
$\sim$ 60 \citep[e.g.,][]{Sheffer2007}. 
$T_{12}$ and $T_{13}$ are the observed peak intensity for CH$_3$OH and $^{13}$CH$_3$OH lines.
Table 2 gives the estimated CH$_3$OH  $\tau_{12}$ for the compact ridge and the outflow, and we 
see that they are mostly $<$ 10. 
We also estimate an upper limit of $\tau_{12}$ $\sim$ 15 for those methanol lines with E$_u$ $>$ 370 K
using the three times of the the band 1a rms (rms: T$_A$$^*$$=$0.065 K) as the upper limit
for all undetected $^{13}$CH$_3$OH emissions.

\begin{table}
\caption{Detected $^{13}$CH$_3$OH counterparts}
\centering
\begin{tabular}{cccccc}
\hline
$^{13}$CH$_3$OH & $\nu$ & E$_u$ & T$_{A}$$^*$ & $\tau_{12}$ & $\tau_{12}$ \\
(J$_{K,{\rm{v}}_t}$) & (MHz) & (K) & (K) & (cr) & (outflow)\\
\hline
14$_{-1,0}$$\rightarrow$13$_{0,0}$ E & 507380.068 & 243.11 & 0.58 & 1.5 & 12.9\\
16$_{-4,0}$$\rightarrow$16$_{-3,0}$ E & 528323.505 & 396.25  & 0.28 &16.9 & 4.4\\
15$_{-4,0}$$\rightarrow$15$_{-3,0}$ E & 528471.541 & 360.05  & 0.34 &7.8 & 4.2\\
14$_{-4,0}$$\rightarrow$14$_{-3,0}$ E & 528606.922 & 326.10  & 0.30 &7.2 & 6.3\\
13$_{-4,0}$$\rightarrow$13$_{-3,0}$ E & 528730.329 & 294.41  & 0.41 &8.7 & 4.8\\
12$_{-4,0}$$\rightarrow$12$_{-3,0}$ E & 528842.385 & 264.98  & 0.57 &6.1 & 6.5\\
11$_{-4,0}$$\rightarrow$11$_{-3,0}$ E & 528943.663 & 237.82  & 0.60 &5.6 & 6.6\\
9$_{-4,0}$$\rightarrow$9$_{-3,0}$ E & 529115.969 & 190.27  & 0.70 &5.7 & 7.3\\
8$_{-4,0}$$\rightarrow$8$_{-3,0}$ E & 529187.882 & 169.89  & 0.70 &4.5 & 9.7\\   
5$_{-4,0}$$\rightarrow$5$_{-3,0}$ E & 529351.258 & 122.33  & 0.62 &5.3 & 8.2\\   
4$_{-4,0}$$\rightarrow$4$_{-3,0}$ E & 529389.156 & 111.00 & 0.65 &6.4 & 3.5\\    
\hline
\end{tabular}
\end{table}

\section{Population diagrams of the isolated transitions}

Population diagrams are one traditional, but effective, way to display
the relative populations of different energy states, generally obtained from a set of observed transitions
\citep[][]{Goldsmith1999}.
This is a useful tool to study the physical conditions in the gas, 
as the relative populations of different states change as a function of 
kinetic temperature and density. 
A population diagram plots the upper state energy E$_u$ as the x axis and
the logarithm of N$_u$/g$_u$ as the y axis, 
where N$_u$ and g$_u$ are the column density and statistical weight  
in the upper state, respectively. 
N$_u$ can be derived from the observed integrated line
intensity of the associated transition 
with the eq. (9) in \citet{Goldsmith1999}.
Therefore, if all energy levels are thermalized with single kinetic temperature (LTE),
the population diagram should be a straight line with the slope and y-intercept presenting
the kinetic temperature and total column density.

\begin{figure*}
  \sidecaption
     \includegraphics[width=12cm]{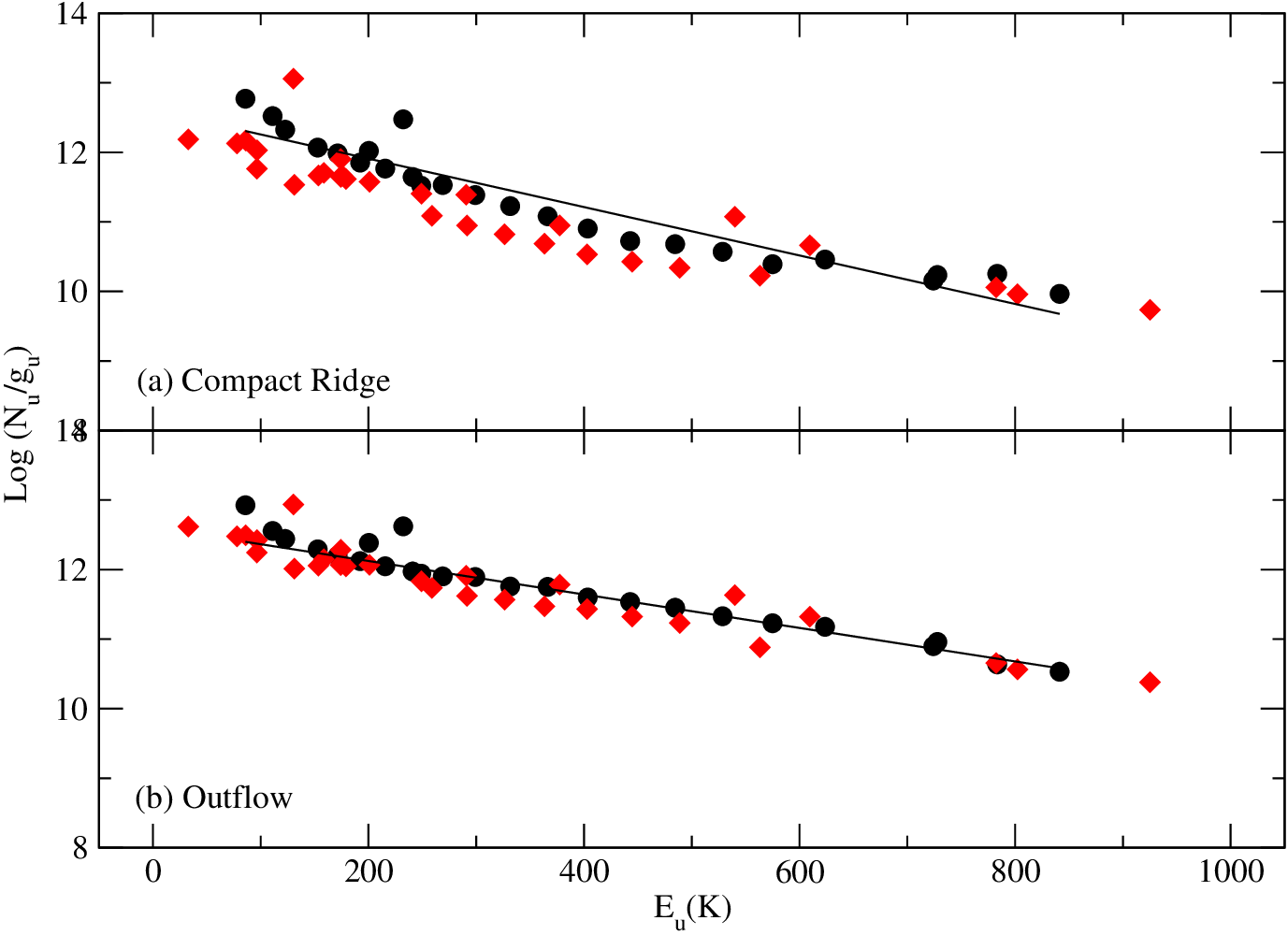}
   \caption{
     Population diagrams of the 524 GHz methanol band (in black) 
     and the other 30 isolated E-type methanol transitions detected 
     throughout the rest of 1a (in red), for (a) the compact ridge and (b) the outflow, respectively.
     The fitted straight lines from Fig.~\ref{fig3} are also plotted for comparison. 
     }
         \label{whole1a}
\end{figure*}

Fig.~\ref{fig3} shows the population diagrams, without correcting for possible beam dilution
and opacity, for the detected isolated 33 transitions of E-type v$_t$$=$0 methanol and their $^{13}$CH$_3$OH
counterparts for the  compact ridge (upper panel) and the outflow (lower panel), respectively.
The straight lines shown in this figure are the slopes determined from least squares fitting with a linear function.
We do not include the HIFI band 4b transitions in the fitting as they are observed with different beam sizes. 
This figure shows that a curvature in the methanol population diagram is clearly seen for the compact ridge,
which will be further examined in \S 5.
However, there are not enough $^{13}$CH$_3$OH transitions to confirm
whether this curvature is also shown for $^{13}$CH$_3$OH in compact ridge.

Unlike the curvature seen in the compact ridge data, the populations of the 1a transitions
from the outflow can be satisfactorily represented by a single kinetic temperature in LTE 
of 184 K and 94 K for CH$_3$OH and $^{13}$CH$_3$OH, respectively.
The $^{13}$CH$_3$OH temperature might be a better approximation to the true kinetic temperature in
this system because its emission is optically thin.
However, the level populations derived from transitions within the 1061 GHz methanol band arising from the outflow 
exhibit a higher degree of scatter compared to those from compact ridge. 
This might suggest that the higher spatial resolution observations at 1 THz are less well
coupled to the outflow and perhaps are contaminated by emission from other components
such as the hot core.
It is intriguing that the outflow in Orion KL is consistent with a single gas temperature, 
as outflow shocks are thought to be gas composed of a range of temperatures.
It is likely that the derived temperature is simply showing an average gas temperature
in this system.

In addition, in order to verify whether the curvature in the compact ridge and the single temperature population in the outflow
are still valid for those transitions outside the selected distinctive methanol bands,
we compare the population diagram of the 524 GHz methanol band 
with that from the  30 isolated E-type v$_t$$=$0 transitions in HIFI band 1a. 
Fig.~\ref{whole1a} plots the population diagrams for the 524 GHz methanol band (in black) and the rest 30 lines (in red).
A curvature is still present in compact ridge and all the level populations in the outflow can still be fitted with a straight line.
The fitted straight lines indicating rotation temperatures of 125 and 184 K from Fig.~\ref{fig3} are plotted
for comparison. This figure also shows that the addition of these 30 transitions increases the scatter
in the population diagram for both compact ridge and outflow. 
As these 30 transitions are actually dominated by different Q branches, compared to those in the 524 GHz methanol band,
this scatter is possibly due to a deviation from LTE for populations within different K ladders. 

\section{Probing the thermal structure in the Orion KL compact ridge}

\begin{figure*}
  \sidecaption
     \includegraphics[width=12cm]{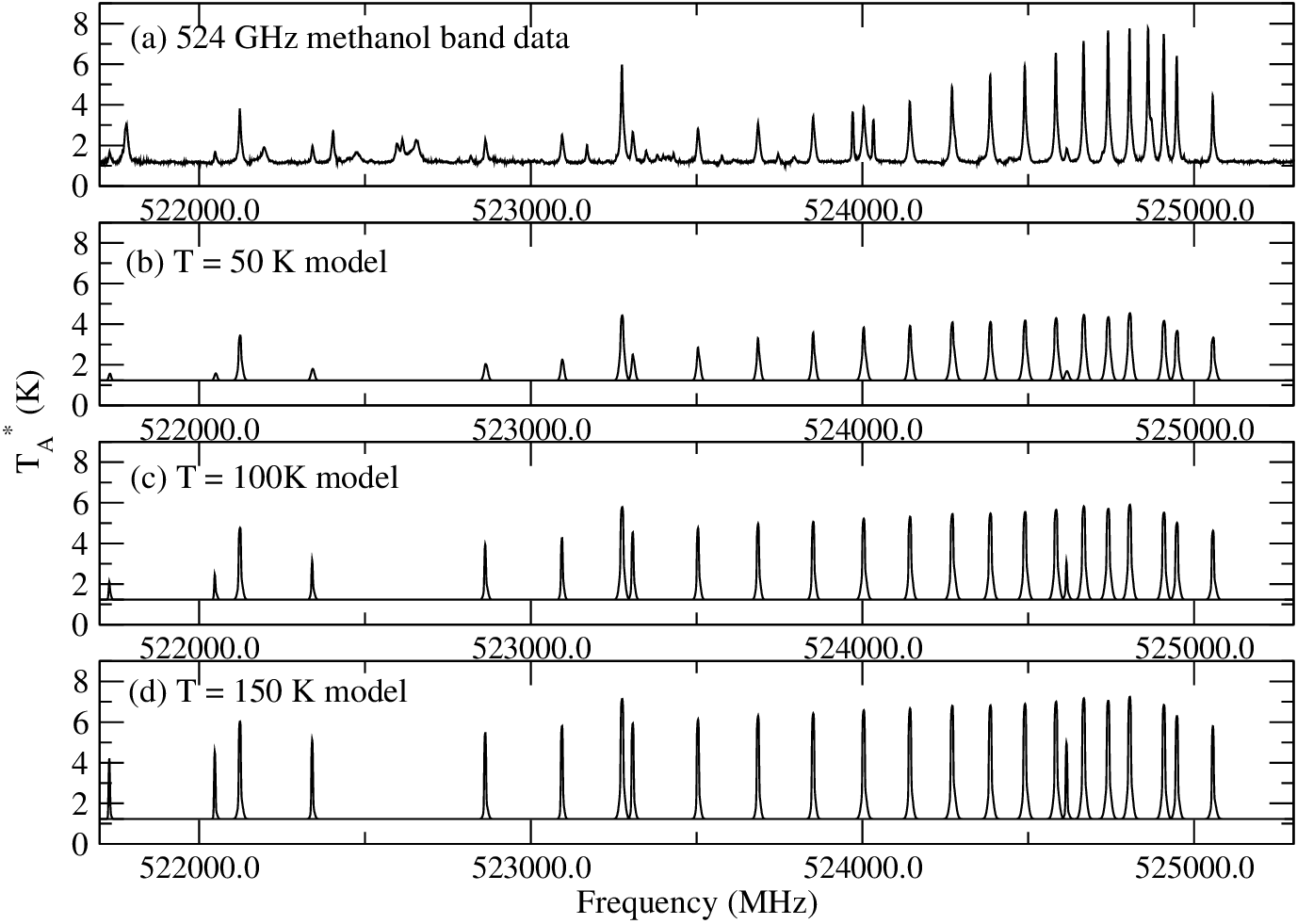}
   \vspace{0.5cm}
   \caption{
     Modeled spectra compared with the those observed in the 524 GHz methanol band (panel a), 
     for models with constant temperature of 50 (panel b), 100 (panel c), and 150 K (panel d), 
     assuming the compact ridge is a spherical clump 5$''$ in radius (r), constant molecular hydrogen number 
     density 10$^7$ cm$^{-3}$, and methanol abundance 10$^{-6}$.
     }
         \label{spconstT}
\end{figure*}

As shown is \S 4, a curvature is present in the methanol
population diagram of compact ridge, which cannot be fitted
well with  LTE excitation at a single temperature (Fig.~\ref{fig3}). 
It has also been suggested in Fig.~\ref{whole1a} that non-LTE excitation exists for methanol in the compact ridge,
at least between transitions with different K ladders.
Indeed, \citet{Leurini2004} has shown that LTE can be a poor assumption for methanol 
even in clouds with high densities. 
However, to date, only transitions with up to upper state energies $\sim$ 350 K have available
collision rates for any non-LTE modeling; while the 524 GHz methanol band includes
upper state energies from 80 to 900 K. 
In addition, according to Fig.~\ref{fig3}, 0 $-$ 350 K is not a wide enough energy range to clearly
distinguish a curve from a straight line in the population diagram.

Nevertheless, despite the currently limited capability for any non-LTE excitation modeling for
all the detected transitions in the 524 GHz methanol band,
it is intriguing that their population diagram shows such a cohesive curvature for compact ridge. 
This curvature might suggest presences of large opacity variations, 
thermal gradients, and/or non-LTE excitation.
Indeed,  the opacity estimation in \S 3.3 (also see Table 2) is not negligible and  
can indeed affect the population diagram slope.
In addition, thermal gradients are another potential contributor to this observed curvature.
It has been suggested that the compact ridge is externally heated due to its very narrow linewidth
and high temperature, 
inconsistent with normal centrally heated star-forming cores \citep[][]{Blake1987}.

\begin{figure}
   \centering
   \includegraphics[width=0.9\columnwidth]{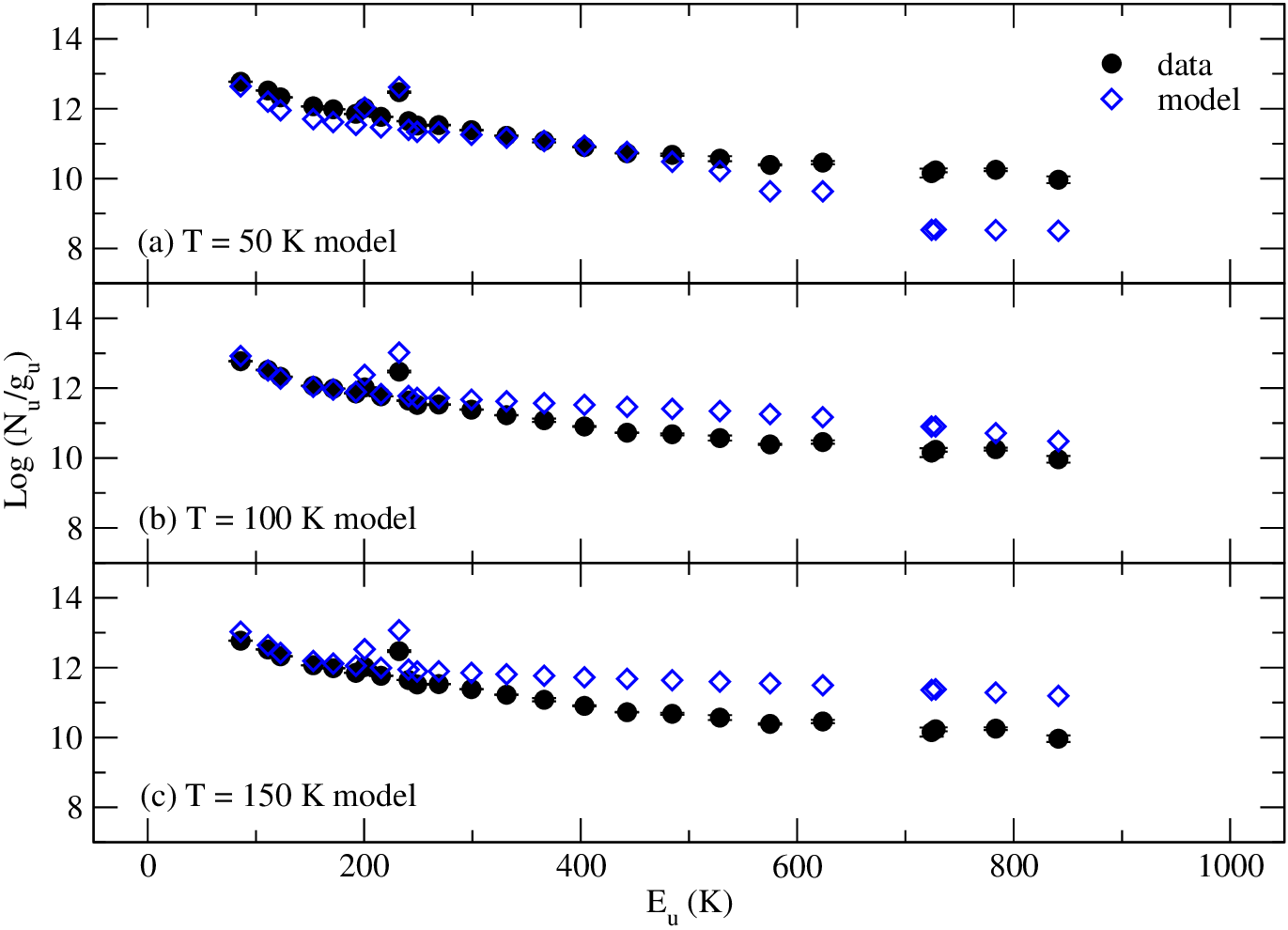}
   \caption{
     Population diagrams of the observed 524 GHz methanol band (in black)
     and those models (in blue) shown in Fig.~\ref{spconstT}.       
     }
         \label{rotdiagconstT}
\end{figure}

In order to investigate how temperature gradients impact the curvature in a population diagram,
we assume the compact ridge to be a spherical clump 5$''$ $-$ 10$''$ in radius (r), having
constant molecular hydrogen number density of 10$^7$ cm$^{-3}$, and a constant
methanol fractional abundance of 10$^{-5}$ $-$ 10$^{-7}$,
as previously suggested \citep[e.g.,][]{Beuther2006,Persson2007}.
We consider various radial temperature profiles, and at each given point in the modeled clump, 
all energy levels are assumed to be populated in LTE with the 
temperature appropriate for that radius. 
With given level populations thus determined, we then use the ray propagation part of 1D RATRAN program \citep[][]{Hogerheijde2000} 
to generate a synthetic spectrum for the 524 GHz methanol band observed with a 40$''$ beam, consistent with the beam size in 
HIFI band 1a.

Fig.~\ref{spconstT} and Fig.~\ref{rotdiagconstT} plot models with constant temperatures, to illustrate how the shape of this
524 GHz methanol band and its population diagram varies with different temperatures. 
Each modeled spectrum is constructed
by combining the modeled compact ridge spectrum with the observed outflow spectrum.
These two plots show that the observed spectrum cannot be explained by  a constant temperature in LTE.

We next explore two approaches with nonuniform temperatures: central heating and external heating. 
First, we examine whether a centrally-heated source can result in the observed spectrum
with a temperature range from a few hundred K in the center of the clump decreasing to 
$~$ 30-50 K in its outer layer, which is consistent with the quiescent gas temperature
along the line of sight \citep[][]{Bergin1994}.  We  allow any methanol abundance variation
within the range 10$^{-5}$ to 10$^{-7}$. 
We were unable to find any models  within these parameter ranges that provided a satisfactory 
match to the observed spectrum. 
The cold layer in the outer part of the clump always results in large opacities which
produce much optically thicker lines than are observed.

\begin{figure*}
  \sidecaption
     \includegraphics[width=12cm]{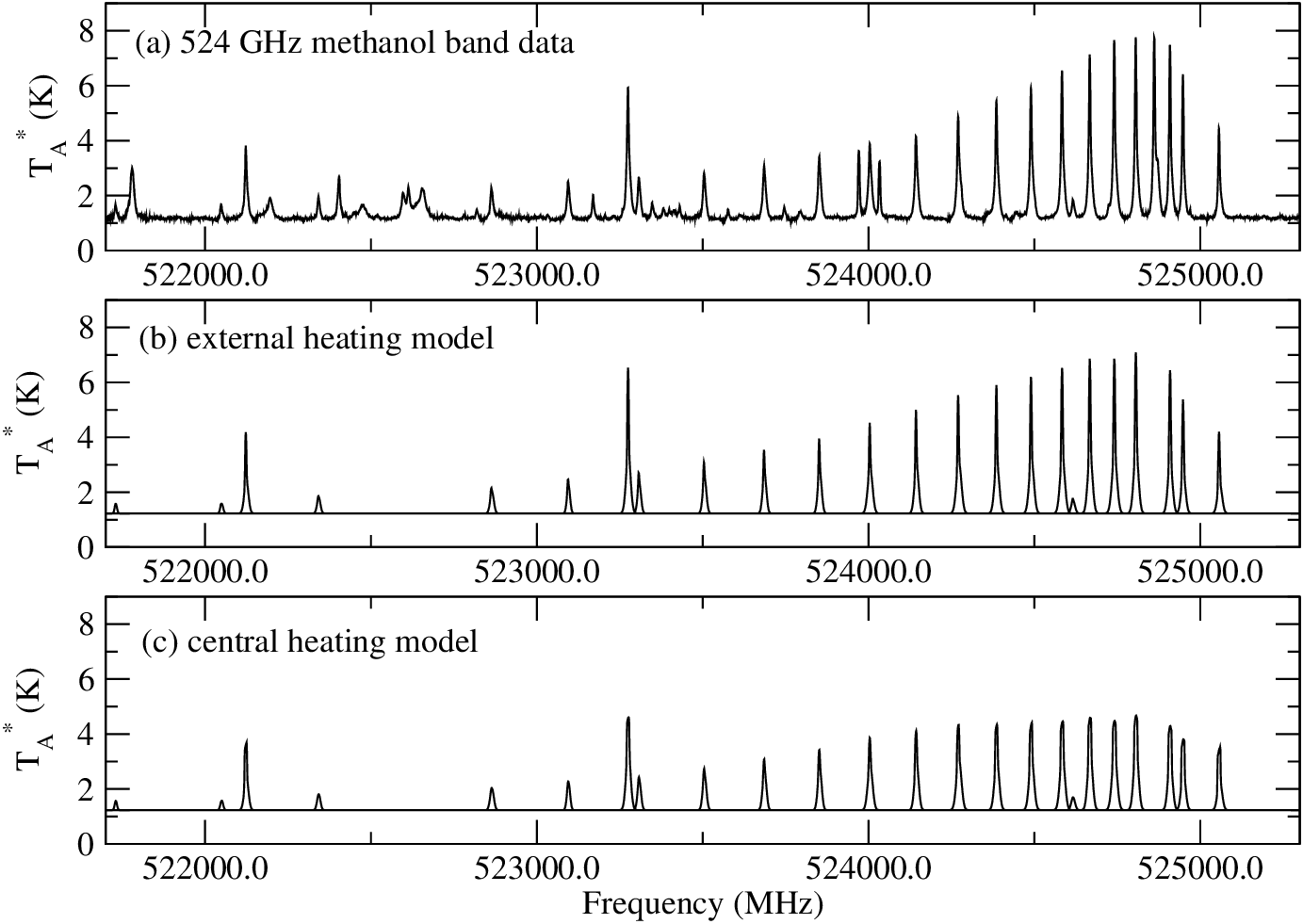}
   \caption{
     (a) The 524 GHz methanol band spectrum.
     (b) The best fit externally heated modeled spectrum.     
     (c) Modeled spectrum of a centrally heated clump with temperature decreasing from 140 K
     in the center to 30 K on the clump surface.
     }
         \label{sp.gradT.eps}
\end{figure*}

\begin{figure*}
     \vspace{2cm}
  \sidecaption
     \includegraphics[width=12cm]{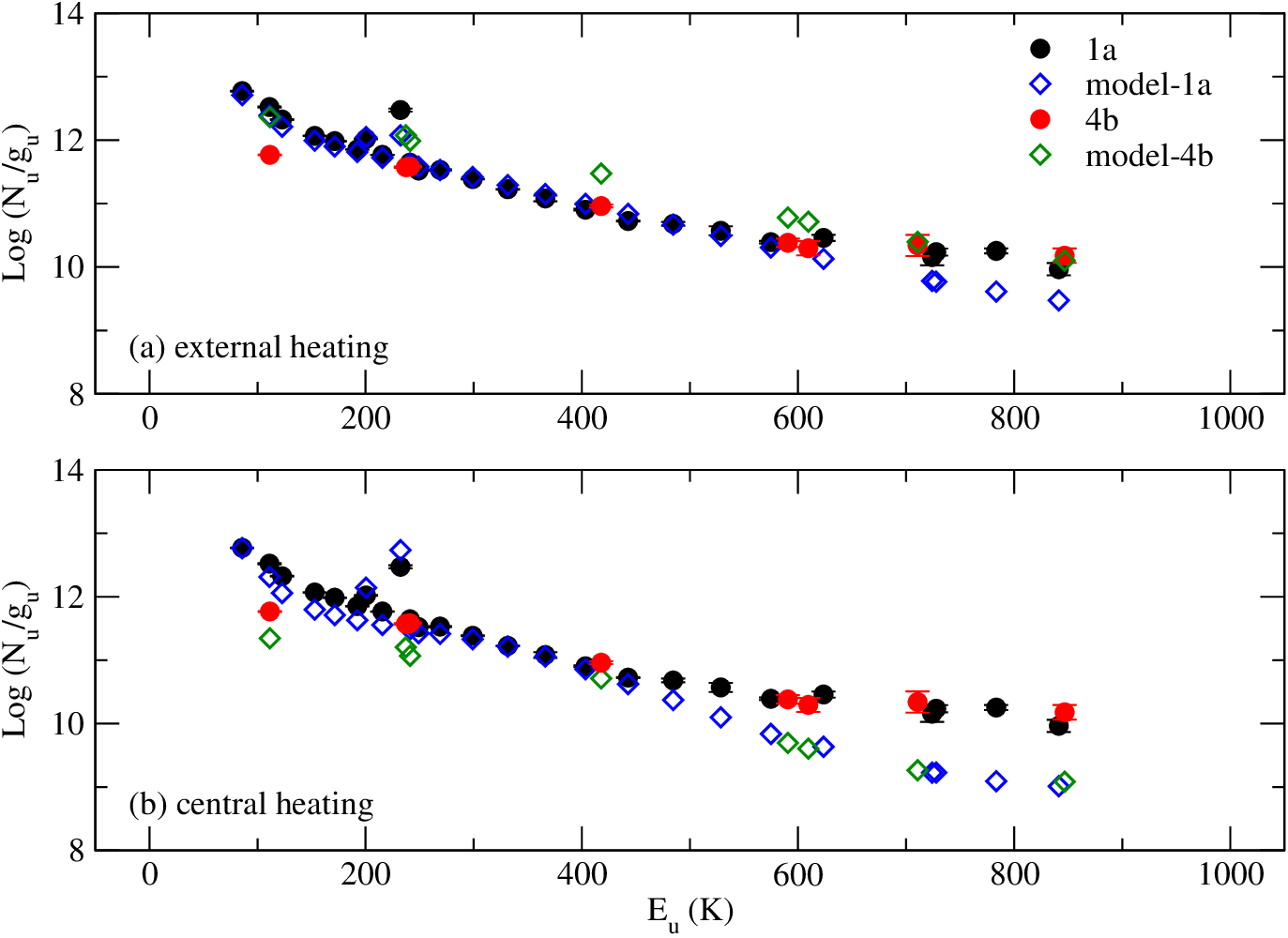}
   \caption{
     (a) Population diagrams for the HIFI band 1a (the 524 GHz methanol band, in black) and HIFI band 4b (
     the 1061 GHz methanol band, in red) methanol bands and
     the best fit externally heated model for  HIFI band 1a transitions (in blue) and HIFI band 4b transitions (in green).
     (b) Population diagrams for the HIFI band 1a (the 524 GHz methanol band, in black) and HIFI band 4b (
     the 1061 GHz methanol band, in red) methanol bands and the 
     centrally heated model (1a in blue and 4b in green) for a clump with temperature decreasing from 140 K
     in the center to 30 K on the clump surface.
     }
         \label{rotdiag.gradT.eps}
\end{figure*}

Second, we considered the possibility of externally-heated clumps and obtained a good fit that
can reproduce the observed spectrum of the 524 GHz methanol band.
The best fit is to a model of the compact ridge as a 7.5$"$ (0.016 pc) spherical clump
with a constant molecular hydrogen number density of 10$^7$ cm$^{-3}$,
a temperature profile with constant 30 K for r$<$5.0$''$, T$\sim$r$^{3.8}$ outside r$>$5.0$''$, and
a constant methanol abundance of 3 $\times$ 10$^{-6}$ and 3 $\times$ 10$^{-7}$ 
in the inner and outer part of the clump. 
This corresponds to a methanol column density of 9.5 $\times$ 10$^{18}$ cm$^{-2}$
and 9.5 $\times$ 10$^{17}$ cm$^{-2}$ in the inner and outer part of the clump, respectively.
This gives a temperature of 120 K on the clump surface.
We stress that the assumed temperature profile in this model is not based on any self--consistent calculation, 
but rather is a way to explore the range of gradients needed to obtain the observed level populations.  
PDR models do show that gas heating can produce such sharp
gradients as modelled here \citep[][]{Kaufman1999} 
and in the future we will examine more physically--motivated solutions.
Fig.~\ref{sp.gradT.eps} plots the 524 GHz methanol band spectrum
and the modeled spectrum
for this best fit externally-heated solution in panel (a) and (b), respectively, and
their associated population diagrams in Fig.~\ref{rotdiag.gradT.eps}a. 
We also apply this model for transitions detected in HIFI band 4b  to generate
its synthetic spectrum appropriate for a 20$''$ beam, and further produce its population diagram shown in
Fig.~\ref{rotdiag.gradT.eps}a. 
Fig.~\ref{sp.gradT.eps}c and Fig.~\ref{rotdiag.gradT.eps}b plot the modeled spectrum and
its associated population diagram for one of the internally heated models we tested, for comparison, in order
to show that the internally heated models can not explain the observed spectrum. 

We conclude that the compact ridge is externally heated with the warm clump surface
in front of the cold gas along the line of sight in LTE.
Although our results show an external heated cloud can indeed result in a curvature in the 
observed population diagram, non-LTE excitation, such as far-infrared pumping, 
is also a possible contributor, which could potentially improve the fit for transitions having higher upper state energies.

\section{Conclusions}

We have examined the methanol emission in HIFI bands 1a and 4b 
from our spectral scan observations toward Orion KL. 
In bands 1a and 4b, 223 and 141 methanol lines are identified.
In band 1a, 112 out of the total 223 lines, and in band 4b, 73 out of the 141 lines  are isolated transitions. 
In order to minimize the confusion due to any blends, abundance variation between E- and A-type methanol,
and optically pumping effects for v$_t$$>$0 transitions, we have focused on 
the isolated E-type v$_t$$=$0 transitions within two methanol bands:
Methanol is known to have numerous transitions from far-infrared to millimeter regime, often including
different symmetric states (A and E), parities, and torsional states, and blends.
This makes the modeling and interpretation of underlying physical conditions difficult. 
But with the approach in this paper -- focusing on one type of isolated methanol transitions --
methanol can be a very useful tool to probe the physical conditions in molecular clouds.
In particular, the 524 GHz methanol band, including 25 isolated E-type v$_t$$=$0 transitions
with upper state energy ranging from 80 to 900 K, provides a valuable set of lines that can be 
easily observed simultaneously and used to examine the temperature and other physical conditions.
Furthermore, by resolving emission from different kinematic components (compact ridge and outflow) for each transition,
we find a curved slope in the population diagram for the methanol emission from the compact ridge.
We suggest that this can be explained by externally heating.
Unlike the compact ridge, the population diagram of the outflow component shows a highly linear relation, suggesting a
single temperature, but which might be the average temperature of the outflowing gas. 
Our study has shown that, with HIFI's broad instantaneous frequency coverage, 
these methanol bands can be very useful for probing the structure of molecular clouds having various physical conditions.

\begin{acknowledgements}
  HIFI has been designed and built by a consortium of institutes and university departments from across 
Europe, Canada and the United States under the leadership of SRON Netherlands Institute for Space
Research, Groningen, The Netherlands and with major contributions from Germany, France and the US. 
Consortium members are: Canada: CSA, U.Waterloo; France: CESR, LAB, LERMA,  IRAM; Germany: 
KOSMA, MPIfR, MPS; Ireland, NUI Maynooth; Italy: ASI, IFSI-INAF, Osservatorio Astrofisico di Arcetri- 
INAF; Netherlands: SRON, TUD; Poland: CAMK, CBK; Spain: Observatorio Astron—mico Nacional (IGN), 
Centro de Astrobiolog'a (CSIC-INTA). Sweden:  Chalmers University of Technology - MC2, RSS \& GARD; 
Onsala Space Observatory; Swedish National Space Board, Stockholm University - Stockholm Observatory; 
Switzerland: ETH Zurich, FHNW; USA: Caltech, JPL, NHSC.   The HEXOS team also is grateful to the HIFI instrument team for building a fantastic instrument.
Support for this work was provided by NASA through an award issued by JPL/Caltech.
\end{acknowledgements}


\end{document}